\documentclass[
    aps,
    prb,
    showpacs,
    reprint,
    superscriptaddress,
    longbibliography,
    floatfix,
    amsmath
]{revtex4-1}

\usepackage{braket}
\usepackage{graphics}
\usepackage{multirow}
\usepackage{dcolumn}
\newcolumntype{.}{D{.}{.}{-1}}

\usepackage{epsfig}
\usepackage{graphicx}
\usepackage{epstopdf}
\usepackage{hyperref}
\usepackage{verbatim}
\usepackage[english]{babel}

\usepackage[utf8]{inputenc}
 
\usepackage{listings}
\usepackage{color}
 
\definecolor{codegreen}{rgb}{0,0.6,0}
\definecolor{codegray}{rgb}{0.5,0.5,0.5}
\definecolor{codepurple}{rgb}{0.58,0,0.82}
\definecolor{backcolour}{rgb}{0.95,0.95,0.92}
 
\lstdefinelanguage{json}
{
    morestring=[b]",
    morestring=[d]',
    string=[s]{"}{"},
    comment=[l]{:\ "},
    morecomment=[l]{:"},
}

\lstdefinestyle{mystyle}{
    backgroundcolor=\color{backcolour},   
    commentstyle=\color{codegreen},
    keywordstyle=\color{magenta},
    numberstyle=\tiny\color{codegray},
    stringstyle=\color{codepurple},
    basicstyle=\footnotesize\ttfamily,
    breakatwhitespace=false,         
    breaklines=true,                 
    captionpos=b,                    
    keepspaces=true,
    numbers=left,                    
    numbersep=2pt,                  
    showspaces=false,                
    showstringspaces=false,
    showtabs=false,                  
    tabsize=2
}
 
\begin{document}

    \title{Data-centric online ecosystem for digital materials science}
    
    \author{Timur Bazhirov}
    \affiliation{Exabyte Inc., San Francisco, California 94103, USA}

    \begin{abstract}

    Materials science is becoming increasingly more reliant on digital data to facilitate progress in the field. Due to an enormous diversity in its scope, breadth, and depth, organizing the data in a standard way to optimize the speed and creative breadth of the resulting research represents a significant challenge. We outline a modular and extensible ecosystem aimed at facilitating research work performed in an accessible, collaborative, and agile manner, without compromising on fidelity, security, and defensibility of the findings. We discuss the critical components of the ecosystem and explain the implementation of data standards and associated tools. We focus initial attention on modeling and simulations from nanoscale and explain how to add support for other domains. Finally, we discuss example applications or the data convention and future outlook.
    
\end{abstract}

    \maketitle
    

    Materials design and discovery is a broad inter-disciplinary research area that presently adopts an ever-increasing number of data-driven approaches with success stories reported in a variety of application fields, including catalysis, carbon capture, hydrogen storage, battery electrolytes, semiconductor, solar energy, magneto-electronics, glasses and optics, adhesives, structural and construction materials among other use cases \cite{jain2013materialsproject, curtarolo2012aflowlib, saal2013openQMD, pizzi2016aiida, nomad}. Such efforts promoted the integration of materials science with information technology at an accelerating pace.
    
    \onecolumngrid
     \begin{figure}[h!]
     \centering
        \includegraphics[width = 0.99\textwidth]{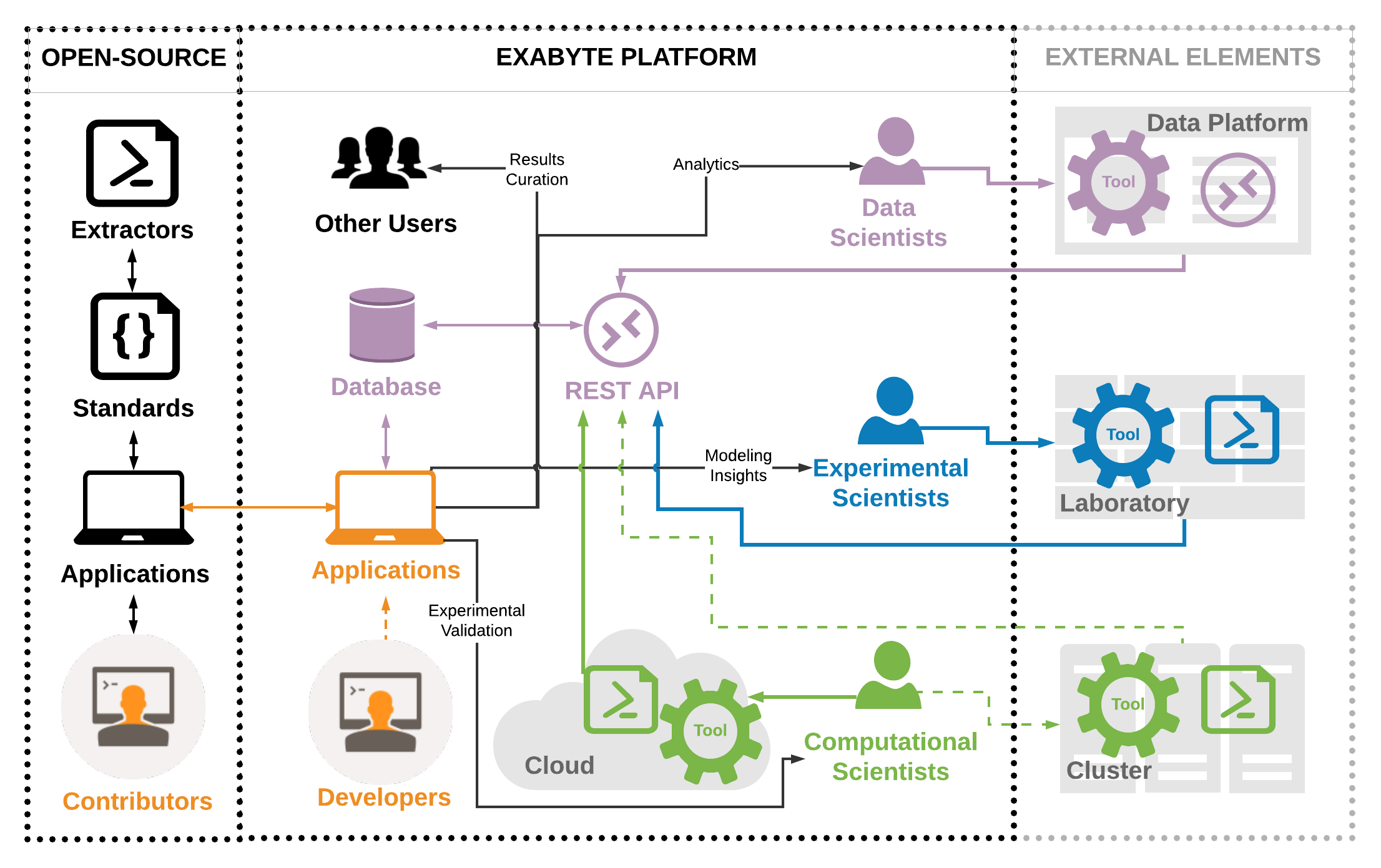}
        \parbox{1.0\textwidth}{
            \caption{
                Schematic representation of the online ecosystem presented in this manuscript. The three main components constitute the "Open-source" panel on the left: "Standards" represent the presently discussed data standards, "Extractors" - associated software tools to convert to the present data convention, and "Applications" represent open-source applications (enabling, for example, the design of new materials geometries) built on top of the data standards. Contributors are shown as well. The "Exabyte Platform" panel demonstrates how the open source components are used within the Exabyte platform. "External elements" panel showcases the elements that are not provided by Exabyte platform but can interface with it. Computational scientists and the logical flow related to their work is shown in green. Similarly, components related to experimental scientists are highlighted in blue, and Data Scientists - in purple. Components and relationships corresponding to Software Development are highlighted in orange. "Cloud" refers to one or multiple cloud computing platforms, "Laboratory" - to an experimental laboratory, "Analytics" - to a set of data analytics tools.
            }
            \label{figure:ecosystem}
        }
    \end{figure}
    \twocolumngrid

     Nevertheless, when compared with the more established computer-aided design and engineering sector, there is still much room for improvement with respect to the organization, standardization and dissemination of data in the digital practices.

    Materials research and development (R\&D) is vast in its complexity and requires a large variety of insights at multiple time- and length scales. Extreme diversity of scales and lack of associated standards and inter-operability increase time-to-insight, limit the agility of the deployed techniques \cite{pizzi2016aiida}, and significantly reduce the impact of digital approaches on the R\&D. Transferability and collaboration are also significantly inhibited. Establishing a standardized and collaborative approach to digital research practices instead will allow to achieve significant improvements.
    

    Here we present the concept of a data-centric online ecosystem allowing to structure and organize the data about materials and chemicals, their properties, and the techniques (workflows) used to extract such properties. The ecosystem is enabled by the associated data convention and is designed with an aim to improve the speed of materials research. The approach is aimed to provide an efficient way to organize a diverse portfolio of research efforts, facilitate inter-disciplinary collaboration, embrace data-centric automated research methodologies, encourage the application of machine intelligence, and allow the broader materials community to contribute their insights to extend the initial prototype presented here.

    
     \begin{figure}[t!]
     \centering
        \includegraphics[width = 0.49\textwidth]{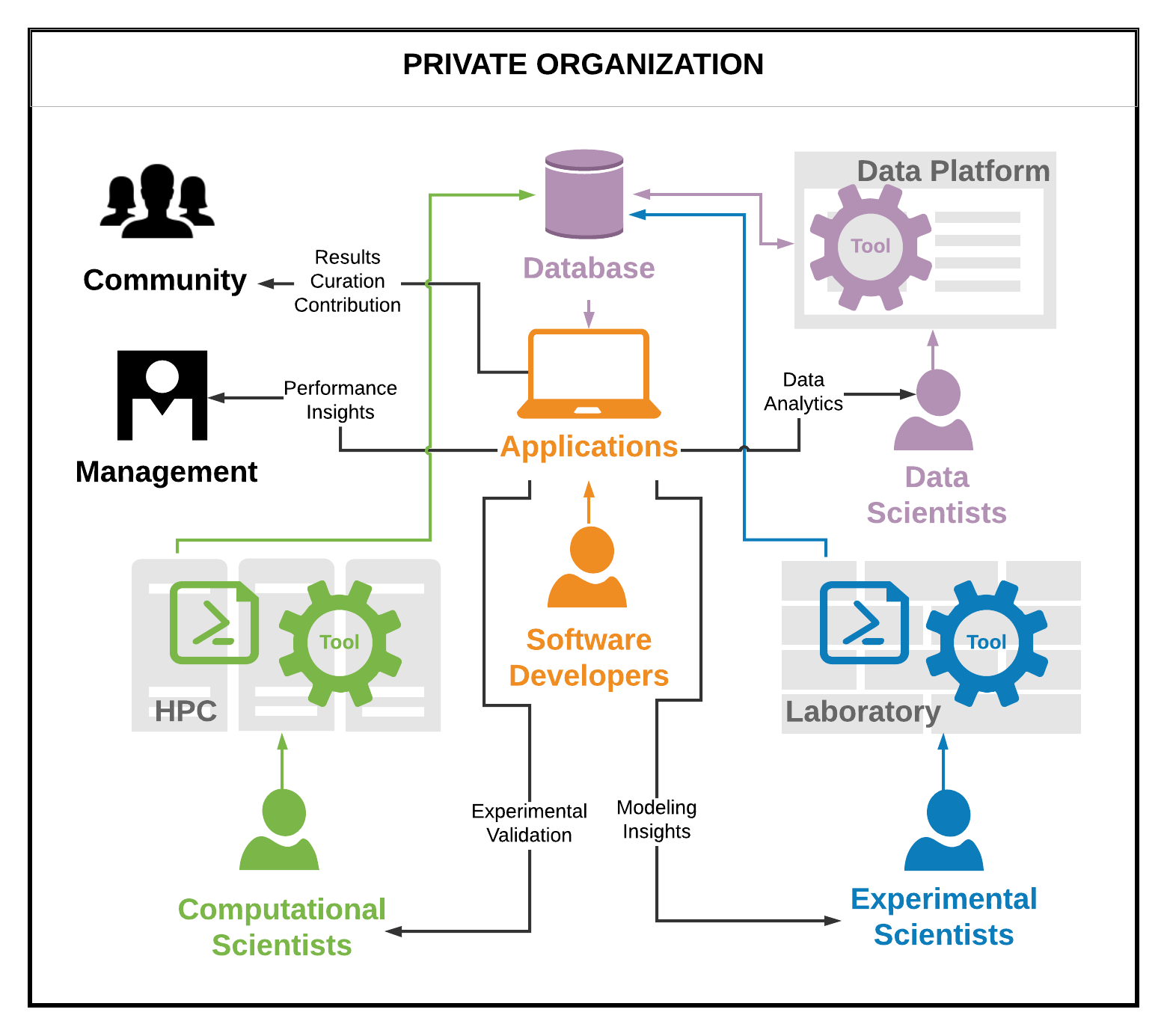}
        \caption{
            Schematic representation of the online ecosystem presented in this manuscript for an example deployment within a private organization. The diagram demonstrates how a secure collaborative environment can be established entirely within a private domain. The open-source parts of the ecosystem presented in Fig. \ref{figure:ecosystem} are used and can be modified to accommodate specific needs of the organization. The legend and icons follow Fig. \ref{figure:ecosystem}. 
        }
        \label{figure:private-ecosystem}
    \end{figure}
    
    
        \textbf{Ecosystem and its Components.} 
        We identify the following main constituents of the ecosystem at the top level, also explained in Fig. \ref{figure:ecosystem}.

        \begin{itemize}
            \item \textbf{Standards}: (or "Data Convention") a set of formats and practices used in structuring the information about materials,
            \item \textbf{Extractors}: tools facilitating the extraction and digitization of materials data and its conversion according to the above Standards,
            \item \textbf{Digitally-enabled equipment}: experimental or computational equipment used during the research activities,
            \item \textbf{Applications}: software utilizing at its core the above standards, including, for example, tools that allow to design new Materials from atoms, or construct sophisticated simulation Workflows.,
            \item \textbf{Contributors}: people that contribute to the development of the above items,
            \item \textbf{Data Repository}: organized source(s) collecting the data and making it accessible online.
        \end{itemize}

        As given in Fig. \ref{figure:ecosystem}, the open-source part of the ecosystem contains the Extractors, Standars, and Applications, and is maintained by Contributors. The latter include the Exabyte platform stuff and any other interested parties willing to participate. In the fiture, we demonstrate the Exabyte platform \cite{exabytePlatform} as an example implementation of the Data Repository. Within the platform a set of (software) developers create applications, using the open-source codebase mentioned above. The applications enable access to data ("Database" and "REST API", or RESTful application programming interface) for a set of platform users. We identify three types of users: computational scientists, experimental scientists and data scientists. The former can utilize cloud resources or external high-performance computing clusters to use modeling tools and feed the data to Database via using Extractors. Similarly, experimental scientists can feed data from their laboratory equipment as given in the figure. Data scientists can access the data using their preferred data platform through REST API. This way all parties collaborate such that modeling experts can get experimental validation, experimentalists can get access to modeling insights, data scientists - conduct analysis, and the broader community - view and curate results.

        \textbf{Data Convention.}
        We identify key Entities and deploy Object-oriented Design principles to construct a set of associated data representation rules. We adopt JavaScript object notation (JSON) as an intuitive, flexible and proven standard for data organization and storage.

        We identify the following example Entities, described in more details further in this manuscript:

        \begin{itemize}
    
            \item \textbf{Material}: holds information about individual materials,
            \item \textbf{Workflow}: contains a logical set of operations, applicable to multiple materials, required to obtain one or more \textit{Properties}
            \item \textbf{Property}: any measurable quantity providing information about another Entity,
            \item \textbf{Job}: a task (experimental or computational) producing Properties for a certain Material-Workflow combination
            
        \end{itemize}

        The above entities apply to both computational and experimental research. We focus on the former in the current writing.
            
        We use "Schema" as a short notation for database schema, a general concept referring to how structured data is stored and organized within a database. We adopt the official public standard for JSON schemas \cite{JSONSchemaDotOrg} providing a vocabulary that allows to **validate** JSON-based documents, and to annotate them with **descriptions**. We include a shortlist of schemas and their corresponding examples in order to demonstrate how the data convention described in this manuscript can be applied in practice. The full list of schemas, examples and the corresponding tools for data validation and usage in software development environments will be made available online soon at the link in Ref. \cite{exabyteESSEGithubRepo}.

        The default JSON schema standard contains a set of primitive types allowed by schema, such as String or Number. We add a set of custom primitive types convenient for storing numerical data and the information specific for materials science considerations. Example additional primitive types include: \textit{axis} - describes the type of data on the axis and its units, used for plotting, \textit{array\_of\_3\_numbers}: - reused to represent a point in 3D space, \textit{scalar}: a datapoint with one numerical value and associated units.
    
        We define the formats and sub-categorize them for each of entities involved. For example, Workflows hold the logic executed during the course of simulations. We further specify the following components of a Workflow:

            \begin{itemize}
                \item \textbf{Subworkflow}: as a set of distinct **units** (elementary calculations) combined together in a flowchart (algorithm), in order to extract one or more properties. In the context of computational studies, a subworkflow must be specific to a particular simulation engine, model and method.
                \item \textbf{Subworkflow Unit}: the most elementary form of information extraction (computation, experimentation) present within a workflow
            \end{itemize}

        We introduce a concept of a "Directory", containing similar formats. Properties Directory, for example, contains the list of properties. We provide some examples below:

            \begin{itemize}
                \item \textbf{Total Energy}: The ground state energy (free energy) of the system.
                \item \textbf{Electronic Density of States}: density of electronic states including partial contributions.
                \item \textbf{Stress Tensor}: 3x3 matrix expressing stresses in spatial dimensions.
            \end{itemize}

        Below is the schema for Total Energy as a representative Property:

\begin{lstlisting}[language=JSON]
{
    "$schema": "http://json-schema.org/draft-04/schema#", 
    "allOf": [
        {
            "required": [
                "name", 
                "units"
            ], 
            "allOf": [
                {
                    "required": [
                        "value"
                    ], 
                    "type": "object", 
                    "properties": {
                        "value": {
                            "type": "number"
                        }
                    }, 
                    "title": "scalar schema"
                }
            ], 
            "properties": {
                "units": {
                    "enum": [
                        "kJ/mol", 
                        "eV", 
                        "J/mol", 
                        "hartree", 
                        "cm-1", 
                        "rydberg", 
                        "eV/atom"
                    ]
                }, 
                "name": {
                    "type": "string"
                }
            }, 
            "title": "energy schema"
        }
    ], 
    "properties": {
        "name": {
            "enum": [
                "total_energy"
            ]
        }
    }, 
    "title": "total energy schema"
}
\end{lstlisting}

        and the corresponding example:

\begin{lstlisting}[language=JSON]
{
    "units": "eV", 
    "name": "total_energy", 
    "value": -123.43573079
}
\end{lstlisting}

        \textbf{Extractors.}
        We maintain a set of tools able to facilitate the conversion into- and from the data standards explained earlier. Example conversion formats include the ones deployed by the Materials Project \cite{jain2013materialsproject}, POSCAR and CIF \cite{kresse1996software, cifFormatHall}, and Quantum ESPRESSO \cite{QE2009mainReference}, for example, among others. The open-source nature of the tools and their simplicity allows for the quick addition of new formats. The tools implemented in python are centrally organized and made available in an online repository (ExPrESS) available at the link in Ref. \cite{exabyteESSEGithub}.

        \textbf{Digitally-enabled equipment.}
        Digitally enabled equipment represents the place where the insights about materials and their properties are collected. It a general concept and could represent a high-performance computing system executing modeling and simulation, or an experimental facility (eg. electron microscope) that generate data in digital formats, or alternatively a data infrastructure platform where data scientists operate.

        \textbf{Applications.}
        We provide a set of tools necessary for other interested parties to adopt the present data convention, similar conceptually to software development kits. We make the schemas and examples and related functions for the validation of new data available online as a Python and JavaScript packages, allowing software developers to incorporate our efforts in their work. We foresee development activity for the web-based applications, and plan to maintain a set of software repositories aimed to help facilitate future work in the space.
        
        \begin{itemize}
    
            \item \textbf{Made.js}: MAterials DEsign in JavaScript, a JavaScript library for atomistic design of materials,
            \item \textbf{Wave.js}: Web-based Atomic ViEwer, another JavaScript library for 3D visualization of material geometries,
            \item \textbf{Materials Designer}: an example JavaScript Application for the design of materials.
            
        \end{itemize}
        
        The above components also constitute the Exabyte platform \cite{exabytePlatform}.

        \textbf{Contributors.}
        The Data Convention allows for the ability to store additional information. This information can, for example, be administrative in nature containing timestamps and log entries, or metadata describing the specifics of the environment where the information is obtained from. Another important use for this is to allow for the contributions from other interested parties. The landscape of materials science is extremely diverse, so that only a collaborative approach to the establishing a universal data convention has the potential to fully cover it. We welcome contributions from other parties in extending the Data Convention, as well as and in developing additional Extractors and Applications based on the open-source content.

        \textbf{Data Repositories}
        It appears natural to have (one or more) source(s) of data aggregation associated with the presently described convention. Data extracted from multiple sources can be accessed through such central repository and shared between multiple parties. We have developed the practices required to assert the security and protection of data, however, these shall be discussed separately outside of the present writing. Exabyte platform \cite{exabytePlatform} represents one such online data repository, however, the ecosystem presented here allows for the creation of multiple private repositories. The latter can function as independent copies of the platform asserting the intellectual property handling rules and data security policies of the corresponding organization(s) (see Fig. \ref{figure:private-ecosystem}).


    \textbf{Collaboration.} 
    Due to the breadth and the associated complexity of materials research enabling efficient collaboration is key to sustained progress in the field. We define efficient collaboration like the one that allows multiple parties to work on the same problem together while avoiding any duplicated efforts and keeping the results accessible to authorized parties only. We incorporate these concepts and their representation in the data convention discussed.
    
    The present ecosystem allows for a flexible set of permissions and ownership scenarios, which in turn facilitate secure collaboration and sharing both within an account (by multiple users), between accounts present in the ecosystem, and with the broader community. A set of curators and automated algorithms is deployed to assert the quality and veracity of data available to the ecosystem users.
    
    As it happened for other fields in the past, efficient exchange of data will enable intelligent decision-making in materials research at an accelerated pace. At present there are multiple concerns related to the intellectual property considerations and security of data. We believe these concerns can be resolved through the application of an appropriate set of access rules and data management policies. Resolving these concerns will facilitate a significant progress of the field at large.


    
        \textbf{Examples.} 
        Discussion of the results arising from the application of the convention presented in this manuscript is outside of the scope of the present writing. We therefore only briefly mention example applications and the related results here. The convention allows for the deployment of materials modeling and simulations in a high-throughput and intuitive manner, with prior applications demonstrated for multiple use cases, including metallic alloys \cite{2016-exabyte-aps-abstract}, electronic properties of semiconductors \cite{2018-exabyte-accessible-CMD, 2018-exabyte-binary-compounds}, and vibrational properties of materials \cite{2018-exabyte-phonon}.
        
        We would like to emphasize here that the fundamental value of the present ecosystem is in its modularity, extensibility and applicability to a variety of scenarios on the vast landscape of materials research. The examples mentioned above serve only as a proof-of-concept and are aimed at demonstrating the capabilities of the software system(s) build upon the present implementation.
    
        \textbf{Outlook.}
        Materials innovation helps solve some of the most pressing societal challenges, however, improving the pace of materials discovery will require us to overcome today’s cost-intensive trial-and-error approach and move to using high-fidelity predictive modeling \cite{jain2013materialsproject, nomad, curtarolo2012aflowlib} and digitize complex, multi-dimensional optimization problems requiring a large variety of characterization data at multiple time- and length scales. Obtaining such information is a time- and cost-intensive process. However, once obtained, it also must be stored and managed in an efficient way. We envision that in the future more and more of materials research will involve digital handling of data and contribute to increasing amounts of materials data on the web. The convention presented here can provide a way to organize this data.
    
        Virtual characterization of materials can be employed to design and optimize materials \textit{in silico}, as a characteristic example. In our view, the present ecosystem and associated tools facilitate the next generation of computer-aided design (CAD) tools and enables advanced R\&D capabilities that unlock the development of new kinds of products in critical industries and has the potential to contribute to the digital transformation of the materials sector at large.
    
        The ideas expressed in the present manuscript build upon the Materials Genome Initiative initiated by the US Department of Energy, and are designed to facilitate collaboration between materials scientists (both experimental and computational, both within industry and academia/government) and computer/data scientists to create, deploy and analyze a set of curated methodologies to rapidly study materials at multiple time- and length scales.

    \bibliography{references}

\end{document}